\documentclass[runningheads]{llncs}
\usepackage{graphicx}
\usepackage{tikz}
\usepackage{comment}
\usepackage{amsmath,amssymb} % define this before the line numbering.
\usepackage{color}
\usepackage{multirow}
\usepackage{booktabs}
\usepackage{mathtools}
\usepackage{pifont}% http://ctan.org/pkg/pifont
\newcommand{\cmark}{\ding{51}}%
\newcommand{\xmark}{\ding{55}}%
\usepackage{soul}
\definecolor{maroon}{cmyk}{0,0.87,0.68,0.32}
\usepackage[accsupp]{axessibility}  % Improves PDF readability for those with disabilities.

\newcommand{\diag}[1]{\mathrm{diag}\left(#1\right)}

\newcommand{\boldone}{{\boldsymbol{1}}}

\newcommand{\boldH}{{\boldsymbol{H}}}
\newcommand{\boldI}{{\boldsymbol{I}}}

\newcommand{\boldK}{{\boldsymbol{K}}}
\newcommand{\boldL}{{\boldsymbol{L}}}

\newcommand{\boldh}{{\boldsymbol{h}}}

\newcommand{\boldp}{{\boldsymbol{p}}}

\newcommand{\boldx}{{\boldsymbol{x}}}
\newcommand{\boldy}{{\boldsymbol{y}}}

\newcommand{\boldalpha}{{\boldsymbol{\alpha}}}

\begin{document}
% \linenumbers
\pagestyle{headings}
\mainmatter
\def\ECCVSubNumber{1856}  % Insert your submission number here

\title{Representation Learning with Information Theory for COVID-19 Detection} % Replace with your title

%Papers are limited to 14 pages, including figures and tables, in the ECCV style. Additional pages containing only cited references are allowed. Please refer to the following files for detailed formatting instructions:
%
%Example submission paper with detailed instructions
%LaTeX Templates (tar): eccv2022AuthorKit.tgz
%LaTeX Templates (zip): eccv2022AuthorKit.zip
%Papers that are not properly anonymized, or do not use the template, or have more than 14 pages (excluding references) will be rejected without review.

%% INITIAL SUBMISSION 
%%\begin{comment}
%\titlerunning{ECCV-22 submission ID \ECCVSubNumber} 
%\authorrunning{ECCV-22 submission ID \ECCVSubNumber} 
%\author{Anonymous ECCV submission}
%\institute{Paper ID \ECCVSubNumber}
%%\end{comment}
%%******************
%with the
%Department of Electronics and Informatics (ETRO), Vrije Universiteit Brussel,
%1050 Brussels, Belgium, and also with the imec, 3001 Leuven, Belgium (e-mail:
%erodrigo@etrovub.be; thdo@etrovub.be; ndeligia@etrovub.be).
%\begin{comment}

%\author{Abel Díaz Berenguer \inst{1}\orcidID{0000-1111-2222-3333} \and
%Tanmoy Mukherjeer\inst{2,3}\orcidID{1111-2222-3333-4444} \and
%Nikos Deligiannis\inst{3}\orcidID{2222--3333-4444-5555} \and
%Hichem Sahli\inst{3}\orcidID{2222--3333-4444-5555}}

\author{Abel Díaz Berenguer \inst{1}\and
Tanmoy Mukherjee\inst{1,2}\and
Matias Bossa \inst{1}\and
Nikos Deligiannis\inst{1,2}\and
Hichem Sahli\inst{1,2}}


\institute{Department of Electronics and Informatics (ETRO), Vrije Universiteit Brussel, Pleinlaan 2,
1050 Brussels, Belgium \and
Interuniversity Microelectronics Centre (IMEC), Kapeldreef 75, 3001 Heverlee, Belgium\\
\email{\{aberengu,tmukherj,mnbossa,ndeligia,hsahli\}@etrovub.be}}
%\end{comment}
%******************
\maketitle

\begin{abstract}
Successful data representation is a fundamental factor in machine learning based medical imaging analysis. Deep Learning (DL) has taken an essential role in robust representation learning. However, the inability of deep models to generalize to unseen data can quickly overfit intricate patterns. Thereby, we can conveniently implement strategies to aid deep models in discovering useful priors from data to learn their intrinsic properties. Our model, which we call a dual role network (DRN), uses a dependency maximization approach based on Least Squared Mutual Information (LSMI). LSMI leverages dependency measures to ensure representation invariance and local smoothness. While prior works have used information theory dependency measures like mutual information, these are known to be computationally expensive due to the density estimation step. In contrast, our proposed DRN with LSMI formulation does not require the density estimation step and can be used as an alternative to approximate mutual information. Experiments on the CT based COVID-19 Detection and COVID-19 Severity Detection benchmarks of the 2nd COV19D competition \cite{AI-MIA_ECCV} demonstrate the effectiveness of our method compared to the baseline method of such competition.
%In this paper, we propose a dependency maximization approach based on Squared Mutual Information (SMI) to achieve 
\end{abstract}

\section{Introduction}
%  
%The outbreak of the novel coronavirus (COVID-19) has resulted in global effort to stop its spread. The infectious nature of the virus has caused considerable cost on the economy and the public health system. 
%To contain the spread and transmission of the virus, rapid testing combined with automatic detection are of utmost importance. 
Reverse transcription polymerase chain reaction (RT-PCR) is a widely used screening method and currently adopted as the standard diagnostic method for suspected COVID-19 cases. Accounting for the high false positives in RT-PCR tests, readily available medical imaging methods, such as chest computed
tomography (CT) have been advocated as a possibility to identify COVID-19 patients \cite{doi:10.1148/radiol.2020200343,doi:10.1148/radiol.2020200463,9079648}. Deep learning (DL) has demonstrated impressive results in COVID-19 detection from CT~\cite{Greenspan_et_al_2020}. However, performance generally depends on the prevalence of confounding lung pathologies within the patient cohort. Diseases such as community-acquired pneumonia or influenza have roughly similar visual manifestations in chest CT volumes, which increases the challenges for accurately detecting COVID-19. In addition, the COVID-19 Positives can have different degrees of infection severity. For example, higher severity is usually consistent with increased regions of multifocal and bilateral ground-glass opacities and consolidations \cite{prokop2020corads}. Detecting different degrees of COVID-19 severity entails learning robust feature representations. This situation motivates researchers to apply strategies to enhance feature representations for effectively discriminating between COVID-19 Positives and COVID-19 Negatives, as well as for detecting the COVID-19 severity.

Tarvainen and Valpola~\cite{tarvainen_et_al_2017meante} suggested that consistency regularization guides the two networks of the mean teacher model to adaptively correct their training errors and attain consistent predictions by improving each other feature representations. Consistency regularization has been widely used to learn meaningful feature representations~\cite{sinha2021consistency,hen2020anatomy,ning2021deep}. For instance, in semi-supervised learning (SSL), consistency regularization based methods have proven useful in improving representation learning by harnessing unlabeled data~\cite{bortsova2019semi,cui2019semi,su2019local,liu2020semi,Seo_et_al_2021,Li_SSL_Seg2021}. However, while successful, consistency based methods ignore the structure across different volumes and might impose incoherent consistency~\cite{DBLP:journals/corr/abs-2103-04813}.
%
%They augmented a supervised loss objective by coupling it with a second term to impose consistency regularization on the predictions from the teacher and student networks. 

Mutual information (MI) based representation learning \cite{hu2017learning,hjelm2019learning} methods have been getting attention due to their ability to learn representations that can be employed for further downstream tasks. Though they have achieved promising results, MI estimation is computationally expensive due to the density estimation step. In this work, we argue that maximizing the mutual information on the latent space of DRN, which uses as input perturbated views of the same image, enhances the representations of intricate patterns to boost the ability of deep models to generalize to unseen data. In particular, we propose to use the least squared mutual information (LSMI) \cite{Sugiyama:2012:DRE:2181148}, a powerful density ratio method that has found applications across various domains in machine learning. Our contributions are:

\begin{itemize}
    \item We propose to use a dependency estimation regularization on the latent space and call our model dual role network (DRN). Our proposal involves a loss function that combines the supervised objective with consistency regularization and LSMI to ensure consistency across predictions and MI maximization on nonlinear projections of the DRN latent space.
    \item We evaluate the proposed model on the COVID-19 Detection and COVID19 Severity Detection tasks.
\end{itemize}

\section{Background and Proposed Method}
\label{sec:background_and_propossed_method}
In this section, we provide an overview of our method. While we are motivated by the COVID-19 Detection and COVID-19 Severity Detection problems, this method can be generalized across different medical imaging tasks.
\begin{figure}
	\centering
	\includegraphics[width=20cm,
	height=5cm, keepaspectratio]{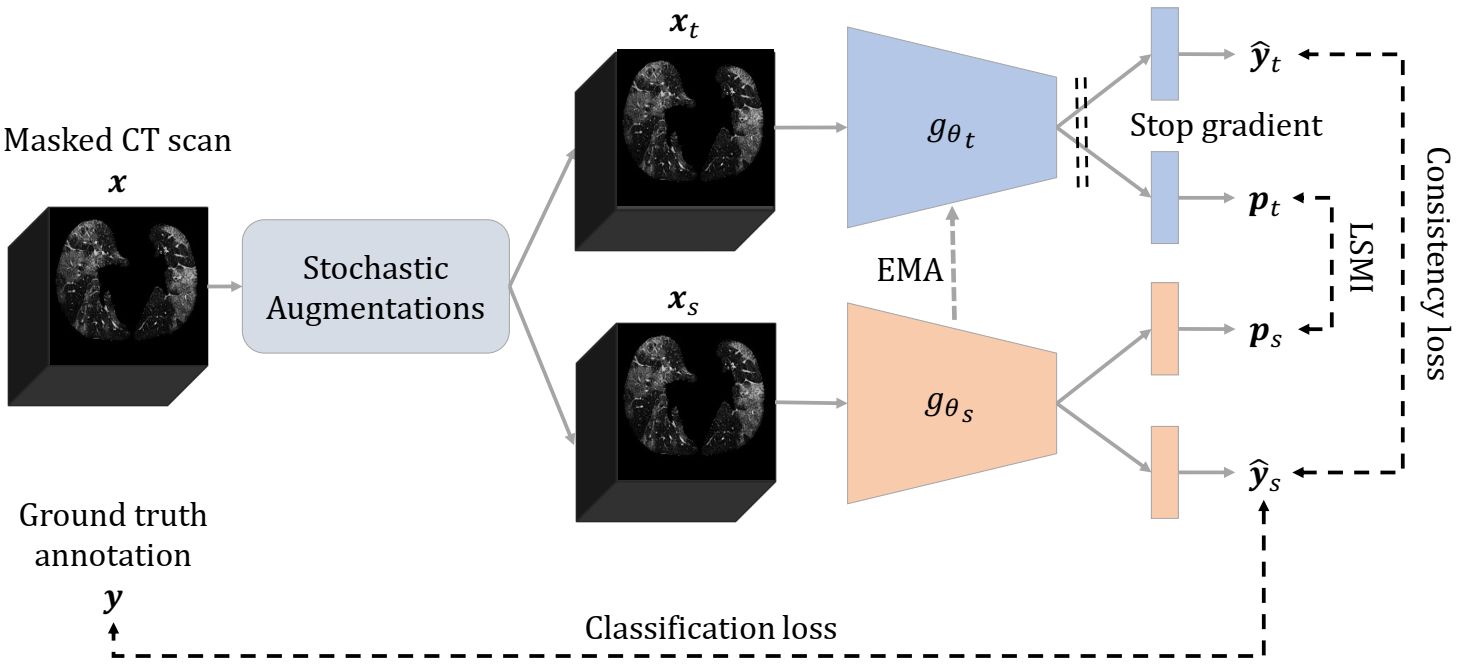}
	\caption{Overview of the proposed method. The masked chest CT scan is stochastically augmented into two views, $\boldx_s$ and $\boldx_t$, repectively. Both transformations pass feed-forward through the dual role networks which output the d-dimensional features projection $\boldp$ and two-dimensional vector $\hat{\boldy}$ for classification. We involve three loss terms to compute a total loss and optimize $g_{\bold{\theta}_s}$. First, the classification loss between ground truth annotation $\boldy$ and $\hat{\boldy}_s$. Second, the consistency loss between $\hat{\boldy}_t$ and $\hat{\boldy}_s$. Third, the LSMI between both networks feature projection $\boldp_t$ and $\boldp_s$, respectively. The total loss is backpropagated to optimize the network $g_{\bold{\theta}_s}$ parameters, and we stop the gradient on the network $g_{\bold{\theta}_t}$ to update its parameters using EMA.} \label{fig1}
\end{figure}

%In semi-supervised learning, we are provided with a small set of labeled data $D_{l} = \{\boldx_l,\boldy_l\}$ where each $\boldx \in \mathbb{R}^{h \times w \times c}$ represents an input masked chest CT volume and $\boldy \in \{0,1\}$ represents the corresponding annotation label indicating if the patient is COVID positive or negative. We also have a set of unlabeled data $D_{u} = \{\boldx_u\}$. We carry a set of perturbations on $\boldx$ to obtain two different views of the input. We aim to obtain two networks, denoted by $g_{\theta_s}$ and $g_{\theta_t}$.

Fig.~\ref{fig1} provides a schema of the proposed method. Given an input masked chest CT scan $\boldx \in \mathbb{R}^{h \times w \times c}$, and its corresponding annotation label $\boldy$, we carry out a set of stochastic perturbations on $\boldx$ to obtain two different views of the input instance, denoted as $\boldx_s$ and $\boldx_t$. Subsequently, we train two networks $g_{\bold{\theta}_s}$ and $g_{\bold{\theta}_t}$ and encourage their output to consistently match. Both the networks share the same backbone architecture (ConvNeXt \cite{liu_et_al_2022convnet} in this work), but these are parameterized by $\theta_s$ and $\theta_t$, respectively. During training, we optimize the parameters $\theta_s$ via backpropagation while $\theta_t$ are updated using exponential moving average (EMA):
\begin{equation}
	\label{eq:eq_ema_update}
	{\theta_t}_\tau=\eta{\theta_t}_{\tau-1}+(1-\eta){\theta_s}_{\tau}
\end{equation}
\noindent where $\eta \in [0,1]$ is the network decay rate coefficient and $\tau$ is the training step. One can notice that the teacher network operates like model ensembling with average exponential decay that have demonstrated usefulness to boost models performance~\cite{polyak1992acceleration}.

To ensure learning robust feature representations, we equip both networks with a projection head on top of the last hidden layer that outputs the latent code used by the classification head. The outcomes of this construction are two. One head outputs $\hat{\boldy}$ for the mainstream predictions concerning the COVID-19 Detection tasks (i.e., COVID-19 Detection and COVID-19 Severity Detection). The second head outputs a $d$-dimensional projection of the feature representation resulting from the last hidden layer, defined as $p \in \mathbb{R}^{d}$. The projection heads are multilayer perceptrons with two linear layers followed by layer normalization and GELU.

To optimize $g_{\theta_s}$, we use the following loss function:
\begin{equation}
    \label{eq_loss}
    J(\theta_s)=H(\hat{\boldy}_s,\boldy)+\lambda\underbrace{d_c(\hat{\boldy}_s,\hat{\boldy}_t)}_{\text{\parbox{4cm}{\centering Consistency\\[-4pt] regularization}}}+\beta\underbrace{d_l(\boldp_s,\boldp_t)}_{\text{LSMI}} 
\end{equation}
%+\beta\underbrace{d_l(\boldp_s,\boldp_t)}_{\text{LSMI}} 
%\text{\parbox{4cm}{\centering Consistency\\[-4pt] regularization}} 
\noindent where $H$ is the cross-entropy loss between the $g_{\theta_s}$ output, $\hat{\boldy}_s$, and the ground truth label $\boldy$. The output of the two classification heads are $\hat{\boldy}_s$ and $\hat{\boldy}_t$. $d_c$ is the mean squared error (MSE) and $d_l$ is the LSMI regularizer between the ouput from projection heads, $\boldp_s$ and $\boldp_t$. For $d_l$ one can also use KL-divergence, and related information theoretic measures. $\lambda$ and $\beta$ are hyper-parameters to control the contribution of the different regularizers. 

%We now focus on the dependency measure we propose to employ in this work. 

%To promote discriminative intermediate feature representations, we equip both networks with a projection head on top of the last hidden layer that outputs the latent code used by the classification head. Thereby, the outcomes of the networks are twofold. One head outputs the logits $\hat{y} \in \mathbb{R}^{2}$ for the mainstream predictions concerning COVID-19 diagnosis task. The second head outputs a $d$-dimensional projection of the feature representation resulting from the last hidden layer, defined as $\hat{p} \in \mathbb{R}^{d}$. The projection head is a Multilayer Perceptron with two linear layers and GELU activation. Inspired by the success of contrastive learning, we argue that latent codes from different views of the same instance should hold similar features statistics with those belonging to the same class distribution. While capable of bringing reliable feature representation, contrastive learning allows regularizing to reduce overfitting. We believe this idea, in general, favors the robustness and generalization capacity of the student and teacher networks. 

%by encouraging its outputs to consistently match with equivalent outputs from a teacher network $g_{\theta_t}$. The student and teacher networks share the same downstream architecture, but these are parameterized by $\theta_s$ and $\theta_t$ respectively. The distinguishing property of the MT 

\subsection{Dependence Measure}
In this work we use {\bf least squared loss mutual information (LSMI)} as a dependence measure. LSMI is an $f$-divergence \cite{10.2307/2984279}, that is a non-negative measure and is zero only if the random variables are independent. Furthermore, to the best of our knowledge, there has not been previous work combining deep models with LSMI as a representation learning metric. To estimate LSMI we take a direct density ratio estimation approach \cite{suzuki2010sufficient}. This leads  \cite{yamada2015cross,yamada2011cross} to the estimator:
\begin{align*}
\widehat{\text{LSMI}}(\{(\boldp_{s_i}, \boldp_{t_i})\}_{i = 1}^n) = \frac{1}{2n} \text{tr}\left( \diag{\widehat{\boldalpha}}\boldK\boldL\right) - \frac{1}{2},
\end{align*}
where $\text{tr}(\cdot)$ is the trace operator,  $\boldK$ is the Gram matrix for $\boldp_s$ and $\boldL$ is the Gram matrix for $\boldp_t$, and $\widehat{\boldalpha}$ is the model parameter written by \cite{suzuki2010sufficient}:
\begin{align*}
\widehat{\boldalpha} = \left(\widehat{\boldH} + \delta \boldI_n\right)^{-1} \widehat{\boldh},~~ \widehat{\boldH} = \frac{1}{n^2} (\boldK\boldK^\top)\circ(\boldL\boldL^\top),~~ \widehat{\boldh} = \frac{1}{n}(\boldK \circ \boldL)\boldone_n.
\end{align*}
%where $\diag{\boldalpha} \in \mathbbR^{n \times n}$ is the diagonal matrix whose diagonal elements are $\boldalpha$.  
\noindent where $\delta$ is a regularizer, and we use the Gaussian kernel: \begin{align*}
\boldK_{ij} \!\!=\! \exp\left(-\frac{\|\boldp_{s_i} - \boldp_{s_j}\|_2^2}{2\sigma_{p_s}^2}\right), ~\boldL_{ij} \!\!=\! \exp\left(-\frac{\|\boldp_{t_i} - \boldp_{t_j}\|_2^2}{2\sigma_{p_t}^2}\right),
\end{align*}
where $\sigma_{p_s} > 0$ and $\sigma_{p_t} > 0$ are the Gaussian width. 
%%The band-width of the kernel function can be optimized using cross-validation \cite{book_hardle}

Most dependency metrics based on $f$-divergence require a \textit{kernel density estimator} (KDE) step to calculate the numerator and denominator, corresponding to the joint probability and the product of its marginals \cite{e15010080}. However, the KDE estimation can be hard to implement and ineffective in practise \cite{book_hardle}. 
To avoid the KDE estimation, an alternative approach involves estimation of the required quantities using $k$-nearest neighbor (KNN) samples \cite{PhysRevE.69.066138}. KNN approach was demonstrated to be better than KDE \cite{PhysRevE.76.026209}. However, it requires one to choose an appropriate $k$, which might not be straightforward.    

LSMI estimator involves simple steps of using linear equations which makes LSMI computationally efficient. 
%One can further develop appropriate CV steps for model selection so that the kernel bandwidth and regularization parameter can be appropriately selected \cite{Suzuki2009MutualIE}. 
Our experiments on using LSMI prove and demonstrate its usefulness as an appropriate measure.

\section{Experimental Results}
\label{sec:results_and_discussion}

\subsection{Datasets Description}
\label{subsec:datasets}

%We evaluated the performance of the proposed method on three different benchmark datasets to validate its effectiveness on different application domains. 

We conducted the experiments to validate the effectiveness of our model on the COVID19-CT-Database (COV19-CT-DB)~\cite{kollias2018deep,kollias2020transparent,kollias2021mia,kollias2020deep,kollias2021miaICCVW,AI-MIA_ECCV}. This dataset is provided to benchmark the models on the 2nd COV19D Competition of the ECCV2022 Workshop: AI-enabled Medical Image Analysis – Digital Pathology \& Radiology/COVID19.

\textbf{COVID-19 Detection Benchmark.} The training set involves $1992$ chest Computer Tomography (CT) scans: $882$ are COVID-19 Positives cases, and $1110$ are COVID-19 Negatives cases. The validation set has a total of $504$ CT scans: $215$ COVID-19 Positives, and $289$ COVID-19 Negatives cases, respectively. Experienced radiologists made the annotations of the CT scans with about $98\%$ degree of agreement~\cite{kollias2021mia,kollias2021miaICCVW,AI-MIA_ECCV}, which makes the COV19-CT-DB appealing to validate our methods because it was not built-up using RT-PCR results as ground truth labels. Further information on the dataset is available in \cite{AI-MIA_ECCV}.

\textbf{COVID-19 Severity Detection Benchmark.} The training set for severity detection comprises $258$ CT scans: $85$ are Mild severity, $62$ are Moderate severity, $83$ are Severe severity, and $26$ are Critical severity. The validation set has a total of $59$ CT scans: with $22$ Mild severity, $10$  Moderate severity, $22$ Severe severity, and $5$ as Critical severity. For more information about this dataset description, readers are referred to \cite{AI-MIA_ECCV}.

%To validate the performance of the proposed model for SSL on the COV19-CT-DB, we trained using $20\%$ labeled and $80\%$ unlabeled data instances.
	
In addition to the data provided for the 2nd COV19D Competition, \textbf{we used the publicly available}\footnote{\url{https://stoic2021.grand-challenge.org/}} \textbf{STOIC dataset to pretrain the backbone used in our model}. The STOIC dataset consists of $2000$ CT scans randomly selected from a cohort of $10000$ subjects suspected of being COVID-19 Positives. The PCR of these $2000$ subjects is provided, enabling us with $2000$ COVID-19 Positives and $2000$ COVID-19 Negatives subjects. We randomly split the data with stratification into training and validation. Finally, to pretrain our ConvNeXt backbone, we employed $930$ COVID-19 Positives and $870$ COVID-19 Negatives subjects in training, with $121$ COVID-19 Positives and $79$ COVID-19 Negatives used for validation.

%In this scenario, we also evaluate our proposal trained semi-supervised using only $20\%$ of labeled and $80\%$ of unlabeled data instances and fully supervised.
%
%%. For CIFAR-10, we use 50,000 unlabeled training images and 10,000 testing
%%images. For SVHN, we use 73,357 unlabeled training images and 26,032 test images. We present
%%SelfMatch results using the average value from three evaluation runs with different random seeds.
\vspace{-5pt}

\subsection{Implementation and Training Details}
\label{subsec:implementation_and_training_details}

We implemented our proposal using Pytorch\footnote{\url{https://pytorch.org/}} upon the PyTorch Image Models~\cite{rw2019timm}. Altougth our method is backbone agnostic, in this implementation we adopt ConvNeXt~\cite{liu_et_al_2022convnet}. To cope with the CT scans for COVID-19 Detection and severity, we extended the ConvNeXt \cite{liu_et_al_2022convnet} to 3D ConvNeXt to harness the spatial information within adjacent slices. Besides, we inflated the weights from the 2D ConvNeXt-B pre-trained on ImageNet to initialize the learning parameters. We also employed the  U-net~\cite{hofmanninger2020automatic} to extract the lung masks from the CT scans to remove non-lung information. Hounsfield units were clipped to the intensity window $[-1150, 350]$ and normalized to the range $[0, 1]$. The masked CT scans were subsequently rezized to $224 \times 224 \times 112$. 

%Unless otherwise stated, we used the described hyper-parameters for all experimental results reported in this work when applicable according to the experiment. 

We trained the models with AdamW during $300$ epoch and early stoping equal $100$ epochs. Following \cite{liu_et_al_2022convnet}, we also used layer scale with initial value of $1e-6$. We linearly ramped up the learning rate during $20$ epochs to $4e-5$, subsequently cosine decaying schedule is used. The weight decay is also scheduled to go from $2e-5$ to $2e-2$ following a cosine schedule. The EMA decay, $\eta$, is equal to $0.9998$. We also used stochastic drop lenght equal to $0.4$. Besides, we employed label smoothing equal to $0.4$. The consistency $\lambda$ and constrastive $\beta$ costs were set equal to $0.1$ and $0.5$, with a linear ramp up during the first $30$ epochs. To find these values we used the Tree-structured Parzen Estimator algorithms \cite{bergstra2013making,bergstra2011algorithms} from Optuna\footnote{\url{https://optuna.readthedocs.io/en/stable/index.html}} \cite{akiba2019optuna}. The values for $\lambda$ and $\beta$ proved to have high impact on the performance of the proposed model. As a general rule, we found that good combinations are with $\beta \le \lambda$, $0 < \beta \le 0.3$, and  $0.3 \le \lambda \le 0.6$. The dimension, $d$, of the projection vectors $\boldp_s$ and $\boldp_t$ is $256$. We used Monai\footnote{\url{https://monai.io/}} to apply images transformations and the stochastic augmentations cosisting of Gaussian smoothness, constrast adjustment, random zooming and intensity shifting. 

\subsection{COVID-19 Detection}

\begin{table}
	\centering
	\caption{The effects of various dependency measures (KL, InfoMax, Jensen–Shannon divergence (JSD), and LSMI) on the performance of our DRN with consistency regularization for COVID-19 Detection. We report the macro F1-score on the validation set of the COV19-CT-DB Detection Benchmark.}\label{tab:Supervised Detection task}
	\centering
	\resizebox{\textwidth}{!}{\begin{tabular}{c|c|c|cccc|c} 
			\toprule
			\multirow{2}{*}{Model}   & \begin{tabular}[c]{@{}c@{}}Classification \\loss\end{tabular} & \begin{tabular}[c]{@{}c@{}}Consitency \\loss ($d_c$)\end{tabular} & \multicolumn{4}{c|}{\begin{tabular}[c]{@{}c@{}}Dependency \\measure ($d_l$)\end{tabular}} &\multirow{2}{*}{\begin{tabular}[c]{@{}c@{}}Macro \\F1-score\end{tabular}}\\ 
			\cline{3-7}
			&CE & MSE & InfoMax & KL   &LSMI  &  JSD \\ 
			\hline
			DRN-MSE-InfoMax & \cmark & \cmark & \cmark   & \xmark  & \xmark & \xmark&$0.86$  \\ 
			%\hline
			DRN-MSE-KL& \cmark  & \cmark & \xmark  & \cmark & \xmark&\xmark & $0.88$ \\ 
			%\hline
			DRN-MSE-JSD& \cmark  & \cmark & \xmark  & \xmark & \cmark & \xmark& $0.89$ \\ 
			%\hline
			DRN-MSE-LSMI & \cmark  & \cmark & \xmark  & \xmark &\xmark & \cmark& $0.90$  \\
			\hline
	\end{tabular}}
\end{table}

We summarize the results of our DRN with consistency regularization and various information theory loss terms in Table~\ref{tab:Supervised Detection task}. We use KL, InfoMax, JSD, and LSMI to assess their impact on the effectiveness of our model in coping with the COVID-19 Detection task. One can notice that our DRN combining consistency regularization and MI maximization is favorable to boosting the performance of deep models in COVID-19 Detection. Furthermore, LSMI regularization (DRN-MSE-LSMI) yielded the best Macro F1 score compared to InfoMax, KL and JSD. All the variants summarized in Table~\ref{tab:Supervised Detection task} outperformed the baseline of the COVID-19 Detection benchmark \cite{AI-MIA_ECCV} whose Macro F1 score was reported equal to $0.77$. Our outcomes support that contrastive learning can be considered as performing maximization of MI to learn representations of data properties beneficial for the mainstream task. The results also suggest that LSMI, used as a representation learning metric, has appealing advantages over other $f$-divergences to reduce the discrepancy between two probability distributions with similar underlying factors while keeping their ``relevant'' information to improve deep models' discriminative ability.

\subsection{COVID-19 Severity Detection}

In Table~\ref{tab:Supervised Severity_Detection} we report the results of our DRN with consistency regularization and different dependency measures, i.e., KL, InfoMax, JSD, and LSMI, to evaluate their impact on the performance of COVID-19 Severity Detection. Similar to the trend exhibited by our DRN in the above section, combining the consistency regularization and the LSMI (DRN-MSE-LSMI) for MI maximization achieved the best Macro F1 score on the COVID-19 Severity Detection task. Our DRN also attained superior results compared to the Macro F1 score of $0.63$ corresponding to the baseline for COVID-19 Severity Detection benchmark \cite{AI-MIA_ECCV}. The experimental results on the COVID-19 Severity Detection task also support the advantages of using LSMI for deep representation learning based on information theory.

\begin{table}
	\centering
	\caption{The effects of various dependency measures (KL, InfoMax, Jensen–Shannon divergence (JSD), and LSMI) on the performance of our DRN with consistency regularization for COVID-19 Detection. We report the macro F1-score on the validation set of the COV19-CT-DB Severity Detection Benchmark.}\label{tab:Supervised Severity_Detection}
	\centering
	\resizebox{\textwidth}{!}{\begin{tabular}{c|c|c|cccc|c} 
			\toprule
			\multirow{2}{*}{Model}   & \begin{tabular}[c]{@{}c@{}}Classification \\loss\end{tabular} & \begin{tabular}[c]{@{}c@{}}Consitency \\loss ($d_c$)\end{tabular} & \multicolumn{4}{c|}{\begin{tabular}[c]{@{}c@{}}Dependency \\measure ($d_l$)\end{tabular}} &\multirow{2}{*}{\begin{tabular}[c]{@{}c@{}}Macro \\F1-score\end{tabular}}\\ 
			\cline{3-7}
			&CE & MSE & InfoMax & KL   &LSMI  &  JSD \\ 
			\hline
			DRN-MSE-InfoMax & \cmark & \cmark & \cmark   & \xmark  & \xmark & \xmark &$0.69$ \\ 
			%\hline
			DRN-MSE-KL& \cmark  & \cmark & \xmark  & \cmark & \xmark & \xmark &$0.71$ \\ 
			%\hline
			DRN-MSE-JSD & \cmark & \cmark &  \xmark  & \xmark  & \xmark & \cmark & $0.72$ \\ 
			%\hline
			DRN-MSE-LSMI & \cmark  & \cmark & \xmark  & \xmark & \xmark & \cmark & $0.73$  \\
			\hline
	\end{tabular}}
\end{table}
\section{Conclusions}
\label{sec:conclussions}
We presented a representation learning method using information theory for COVID-19 Detection and COVID-19 Severity Detection. The proposed estimator is non-parametric and does not require density estimation, unlike traditional MI based methods.
Experiments demonstrate the effectiveness of our proposal. 
We found that DRN with Squared Mutual Information for enhancing representation learning encourages the dual networks to achieve adequate results. Our future work will study the effects of our choices for stochastic augmentations.
%
%We presented a representation learning method using information theory for COVID-19 Detectin and COVID-19 Severity Detection. The proposed estimator is non-parametric and does not require density estimation unlike traditional MI based methods.
%Experiments demonstrate the effectiveness of our proposal for supervised, but also semi-supervised learning. 
%We found that dual role networks with Squared Mutual Information for enhancing contrasting learning encourage the dual networks to achieve adequate results by ensuring invariant representation of different realization of the same image. Our future work will study the effects of our coices for stochastic augmentations. 

%\subsubsection{Reproducibility statement} The code corresponding to the implementation of all models studied in this work and their weights will be available in an online repository. We also make available the train, validation, and test sets used for the experiments on the STOIC dataset.

\bibliographystyle{splncs04}
\bibliography{egbib,refs_Challenge_ECCV2022}
\end{document}